\documentclass[fleqn,usenatbib]{mnras}

\usepackage{newtxtext,newtxmath}

\usepackage[T1]{fontenc}

\DeclareRobustCommand{\VAN}[3]{#2}
\let\VANthebibliography\thebibliography
\def\thebibliography{\DeclareRobustCommand{\VAN}[3]{##3}\VANthebibliography}

\usepackage{graphicx}	
\usepackage{amsmath}	

\title[Be star decretion disc formation]{2D hydrodynamical simulations of Be star decretion disc formation through boundary layer effects}

\author[M. Overton et al.]{
Madeline Overton$^{1,2}$\thanks{E-mail: overtm2@unlv.nevada.edu},
Zhaohuan Zhu$^{1,2}$, 
Rebecca G. Martin$^{1,2}$,
and Jiayin Dong$^{3,4}$
\\
$^{1}$Nevada Center for Astrophysics, University of Nevada, Las Vegas, 4505 South Maryland Parkway, Las Vegas, NV 89154, USA\\
$^{2}$Department of Physics and Astronomy, University of Nevada, Las Vegas 4505 South Maryland Parkway, Las Vegas, NV 89154, USA\\
$^{3}$Department of Astronomy, University of Illinois at Urbana-Champaign, Urbana, IL 61801, USA\\
$^{4}$Center for Astrophysical Surveys, National Center for Supercomputing Applications, Urbana, IL 61801, USA
}

\date{Accepted XXX. Received YYY; in original form ZZZ}

\pubyear{\the\year{}}

\begin{document}
\label{firstpage}
\pagerange{\pageref{firstpage}--\pageref{lastpage}}
\maketitle

\begin{abstract}
Be stars are massive main-sequence stars rotating close to their breakup rate. They possess a decretion disc of material built up due to mass loss from the star, however, there is not a consensus to the mechanism responsible for the formation of the disc because of their sub-breakup spin rates. We present the first 2D hydrodynamical simulations of the formation of a Be star decretion disc from a rapidly rotating star due to boundary layer effects that reduce the rotation rate of the disc close to the star. 
In our simulations with a disc aspect ratio of $h/r=0.1$, a decretion disc forms around a star rotating with $80 \%$ of the breakup rate, but fails when rotating at $70 \%$ of the breakup rate. For a thinner disc, a faster stellar spin may be needed to form a dynamically important decretion disc.
We also demonstrate good agreement between 1D and 2D models. Although this work does not consider the presence of magnetic fields and the angular momentum transport is through viscosity, our results robustly show a Be star disc may be built up hydrodynamically through boundary layer effects, and may play an essential role in regulating the stellar spin.
\end{abstract}




\section{Introduction}

Be stars are massive main-sequence (MS) stars that are rotating at a large fraction of their breakup rates and possess a viscous decretion disc built up from mass ejection from the stellar equator \citep{Lee1991, Pringle1991, Okazaki2001vdd,Okazaki2002, Jones2008,Rivinus2013, Rivinus2026}. The breakup spin rate is  $\Omega_{\rm b} = \sqrt{GM_* / R_*^3}$, where $M_*$ is the mass of the star and $R_*$ is the radius of the star.
Distributions of Be star rotation rates tend to peak between $70\%$ and $80\%$ of the breakup rate \citep[e.g.][]{Porter1996, Rivinus2013, Zorecetal2016, Rivinus2026}. 
A Be star may acquire a rotation rate which is a high fraction of its breakup spin rate due to mass transfer during the late stages of a binary companion's stellar evolution \citep{Pols1991, Shao2014, Bodensteiner2020}. Be stars often have an evolved binary companion such as a stripped Helium star \citep{Wang2021, Klement2022a, Klement2025}, a white dwarf, or a neutron star (belonging to the category of Be/X-ray binaries) \citep{Negueruela1998, Reig2011, Marinoetal2025}. Due to mass loss from the binary companion, a significant amount of material can be captured into an accretion disc around a massive MS star (the Be star progenitor). The accretion of this material on to the star causes the MS star to spin up, leaving behind a rapidly rotating star. Alternatively, in the single star origin scenario the Be star acquires rapid rotation over the course of its MS lifetime, for example through angular momentum transport from the inner contracting regions of the star \citep{Ekstr2008, Hastings2020}. While both single and binary formation channels are viable, the relative contribution of the single star channel to the population of Be stars is unknown \citep{Jones2022}. 

The rapid rotation of Be stars must be a key feature in producing a decretion disc. However, the sub-critical rotation is not sufficient to produce a disc that rotates at Keplerian frequency. There is not a consensus to the additional mechanism responsible for decretion. 
Previous literature examining the formation of a decretion disc has remedied mass ejection by sub-critically rotating object by considering a non-zero central torque \citep{Nixon&Pringle2021}. In this scenario, a central source of angular momentum provides the additional support needed to launch the disc. \cite{Nixon&Pringle2020} suggest that this source of angular momentum is small scale magnetic fields which truncate the inner disc. In this case, the inner edge of the disc then rotates at a lower frequency than the star allowing small-scale magnetic flaring events to feed the disc formation at this radius.
However, this requires a finely tuned magnetic field strength \citep{ud-Doulaetal2018}, and, magnetic fields are undetectable in many Be stars \citep[e.g.][]{Neineretal2012}. 

Non-radial pulsations (NRP) are commonly observed in Be stars and may deposit angular momentum at the stellar surface, resulting in increased rotational velocities allowing for decretion \citep{Ando1986, Osaki1986}. There are also cases of NRPs associated with Be star outbursts \citep[e.g.][]{Neiner2020, Carciofi2025}. However this mechanism is only shown to be viable for Be stars with a rotation that is already nearly critical ($90\%$ of breakup and above), and has not been shown to explain Be phenomena at lower rotation rates, which have been observed as low as $40 \%$ of breakup \citep{Cranmer2005}. 

Recent work has shown that purely hydrodynamical models can lead to the formation of a decretion disc considering a non-zero central torque provided by the star. In the boundary layer between the disc and the star \citep{FKRbook2002,Pringle1981}, the rotation rate transitions from the disc's near-Keplerian rotation to the rotation rate of the central body's envelope. \cite{Dongetal2021} performed 2D hydrodynamical simulations showing decretion in the context of circumplanetary discs due to boundary layer effects, when the planet rotation is above some critical rotation rate. These simulations have an initial accretion disc which can become a decretion disc if the planet is rotating rapidly enough. Based on linear theory, \cite{FuHuangYu2023} further explain this effect by demonstrating how vortices transport angular momentum through the boundary layer. Additionally, with one-dimensional models of Be star boundary layers, \cite{Martinetal2025} found steady state solutions for decretion. 
In the boundary layer, the rotation rate monotonically decreases with increasing radius, from the stellar spin rate to the Keplerian rate. This allows material from the sub-critically rotating object to spread outwards.

In this Letter, we present the first 2D hydrodynamical simulations to show that boundary layer effects can form a Be star decretion disc from a rapidly rotating star with no initial disc. 
For a disk aspect ratio of $h_{\rm disc} / r = 0.1$, we find that decretion occurs for a rotation rate that is $0.8 \ \Omega_{\rm b}$, and decretion does not occur for $0.7 \ \Omega_{\rm b}$, or lower. 
The letter is structured as follows. In section~\ref{sec:methods} we outline our numerical setup. In section~\ref{sec:results} we present our results for varying stellar rotation rates in 2D, and a 1D decretion disk model for comparison. Finally, in section~\ref{sec:discussion} we draw our conclusions and discuss the ability of this method to explain some observed behaviors of Be stars and Be star discs.

\section{Numerical methods}
\label{sec:methods}

\subsection{2D simulations}

We use the {\sc Athena++} \citep{Stoneetal2020athenapp} hydrodynamics code to perform two-dimensional, axisymmetric simulations encompassing the outer layers of the star and the inner disc. Our numerical setup closely follows that of \cite{Dongetal2021} except we do not include an initial accretion disc, and instead the initial density at the outer edge of the central object drops exponentially, as described below.
Our simulations use spherical-polar coordinates ($r, \theta, \phi$). The domain consists of $2048$ logarithmically spaced grid-points in the radial direction with a ratio of $1.0011764$ where $r \in [0.9, 10] \ R_*$, and $2048$ uniformly spaced grid-points in the polar angle, where $\theta \in [0, \pi]$. The simulation is azimuthally symmetric.

For the $\theta$ boundaries, we apply the polar-wedge boundary condition, in which the azimuthal velocity changes sign across the poles. For the $\phi$ boundaries we use a periodic boundary condition. The inner radial boundary has a reflecting radial momentum boundary condition, and the hydrostatic equilibrium equation is extrapolated to the ghost cells. There is an outflow boundary condition on the outer radial boundary, and inflow is prohibited by setting the radial velocity in the ghost cell to zero if the radial velocity points inwards. We use $4$ ghost cells to maintain hydrostatic equilibrium at the inner radial boundary where the density of the central object is high.

The gravitational potential of the star is treated as a point mass at the origin since the mass of the stellar envelope outside of the simulation's inner boundary is small. The stellar envelope is initialized between $r = 0.9$ and $1.1 \ R_*$ in non-rotating hydrostatic equilibrium, as described by 
\begin{equation}
    c_s^2 \frac{d \rho}{dr} = -\frac{\rho GM_*}{r^2}, 
\end{equation}
with an initial density profile given by 
\begin{equation}
    \rho(r) = \rho_* \text{exp} \bigg[ \frac{GM_*}{c_s^2 R_*^2} \bigg(\frac{R_*}{r} - 1 \bigg) \bigg]. 
\end{equation}
Additionally, we set $GM_*$, $R_*$, and $\rho_*$ to unity, and we implement a density floor of $10^{-8} \ M_*/R_*^3$ to avoid any grid cells approaching zero density. 
We use an isothermal equation of state, given by $P = c_s^2 \rho$.
There is initially no disc, however the entirety of the simulation domain uses an isothermal equation of state where $c_s$ is a constant. 
Therefore, the disc scale height $h_{\rm disc}$ scales with radius as $ h_{\rm disc} = c_s / \Omega_K \propto r^{3/2}$ and the disc aspect ratio $h_{\rm disc} / r \propto r^{1/2}$, resulting in a flared disc. This is reasonable since Be star discs are thought to be flared such that the disc aspect ratio $h_{\rm disc} / r $ increases with radius \citep{Hanuschik1996}. 

We choose the sound speed $c_s$ such that $h_{\rm disc} / r = 0.1$ at $r=1 \, R_*$. 
Through a variety of methods, Be star discs have been found to possess a wide range of disc aspect ratios, encompassing $h_{\rm disc} / r = 0.1$. Statistics of shell stars suggest disc aspect ratios near $h_{\rm disc} / r = 0.1$; \cite{Porter1996} finds $h_{\rm disc} / r = 0.09$ or a characteristic half opening angle of $5^{\circ}$, and \cite{Hanuschik1996} finds $h_{\rm disc} / r = 0.23$ or a half opening angle of $13^{\circ}$.
An upper limit for the aspect ratio of $\zeta$-Tau was found to be $0.36$ (corresponding to a opening half-angle of $20^{\circ}$) \citep{Quirrenbach1997}. However, this is an upper limit for the outer edge of the disc, and $h_{\rm disc} / r$ may be smaller for the inner regions of the disc. Using polarization data which can probe the inner regions of the disc, \cite{Wood1997} showed the aspect ratio of $\zeta$-Tau is $h_{\rm disc} / r = 0.04$ or a half-opening angle of $2.5^{\circ}$ \citep[see also][]{Carciofi2006,Carciofi2009}. These estimates are derived using the surface temperature of the star. However, in our simulations the disc connects to the upper layers of the star where the temperature may be much higher than it is at the surface. We make the simplifying assumption that the sound speed is constant throughout the upper layers of the star and the disc, but more complete radiation hydrodynamics would be valuable in future work.

The simulations are run in two stages. In stage~1, the kinematic viscosity is set to zero and the angular velocity of the initially non-rotating stellar envelope is gradually increased by adding a constant amount of angular momentum at each timestep until the desired rotation rate at the equator is achieved. 
Angular momentum is added such that the angular frequency is increased by $4 \pi \times 10^{-5}$ rotations per orbit, each timestep, at the equator. In the case the desired angular velocity is exceeded, the angular momentum is damped by a factor of approximately $0.0016$ per timestep, corresponding to an exponential decay with a damping timescale of $100$ orbital periods ($P_{\rm orb}$) at $r = 1 \ R_*$. The rotation is kept at zero at the inner boundary, and at the poles to avoid numerical instabilities at the boundaries. The angular velocity profile in the region $r < 0.96 \ R_*$ is given by a logistic function,
\begin{equation}
    \Omega(r, \ \theta) = \frac{\Omega_* \sin \theta}{1+ \exp [-400 (r-0.93)]} .
\end{equation}
We do not modify the angular velocity for $r \geq 0.96 \ R_*$. While this angular velocity profile is not reminiscent of a realistic stellar interior, it sufficiently produces the desired stellar rotation at the inner edge of the boundary layer for the equatorial region of the star, which is the region of interest for this work.
The rotation rate of the star at the equator, $\Omega_*$, is set to a fraction of the Keplerian orbital frequency at the equator of the star, which is the critical angular velocity for breakup, $\Omega_{\rm b}$. We run simulations with $\Omega_* / \Omega_{\rm b} = 0.7$ and $0.8$, representing rotation rates just below and well within the range for the Be phenomena, and one slowly rotating case with $\Omega / \Omega_{\rm b} = 0.3$. We run stage~1 for a time of $170$ orbits at $r=1 \ R_*$. 

In stage~2 we allow for viscosity by setting the kinematic viscosity 
\begin{equation}
\nu = \alpha \frac{c_s^2 }{ \Omega_K}
\end{equation} 
to a non-zero constant for $r \geq 0.96 \ R_* $. 
Thus, the \cite{SS1973} viscosity $\alpha$ is a decreasing function of radius. We choose the kinematic viscosity such that $\alpha = 0.1$ at $r=1 \ R_*$. We continue forcing the angular velocity interior to $r=0.96 \ R_*$ to the desired value throughout stage~2. However, in stage~2 the $\Omega_* / \Omega_{\rm b} = 0.7$ case requires that angular momentum is added more aggressively, by a factor of $8$ compared to the stage~1 values in order to avoid a discontinuity in $\Omega$ within the stellar envelope. 
Stage~2 is run for an additional time of $340 \ P_{\rm orb}$ after which the simulations have achieved a quasi-steady state in the inner disc region ($r <2 \ R_*$). The outer portions of the disc do not reach a steady state. However since we are primarily interested in the interaction between the stellar envelope and the inner disc, it is sufficient for only the inner portions of the disc to be quasi-steady. 

\subsection{1D Simulations}
To complement the 2D simulations, we also solve  1D vertically integrated equations
\begin{align}
\frac{\partial \Sigma}{\partial t}&=-\frac{1}{R}\frac{\partial (R\Sigma v_{R})}{\partial R}\,\\
\frac{\partial v_{R}}{\partial t}&=-\frac{1}{2}\frac{\partial v_{R}^2}{\partial R}+(\Omega^2-\Omega_{K}^2)R-\frac{1}{\Sigma}\frac{\partial P}{\partial R}\,\label{eq:vrc}\\
\frac{\partial (\Sigma \Omega R^2)}{\partial t}&=-\frac{1}{R}\frac{\partial (\Omega R^3\Sigma v_{R})}{\partial R}+\frac{1}{R}\frac{\partial }{\partial R}\left(R^3\nu\Sigma\frac{\partial \Omega}{\partial R}\right)\,,
\end{align}
using the finite volume method, where $R$ is the cylindrical radius, $\Sigma$ is the disc surface density and $v_{R}$ is the radial velocity. The Strong-Stability Preserving two step Third-order Runge-Kutta time-stepper (SSP-RK3) has been used for the time integration. For each RK3 stage, the change of surface density and the angular momentum due to advection are calculated using the upwind method and piece-wise linear reconstruction with the minmoid limiter. The change of $v_R$ needs special attention. First, we calculate the advection part using the Rusanov/Local Lax-Friedrichs (LLF) flux which includes the characteristic diffusion speed of $v_R$+$\epsilon c_s$. The addition of $\epsilon c_s$ is crucial for the stability of the LLF numerical scheme, since it adds a diffusion term ($(v_R+\epsilon c_s)\Delta R \partial^2 v_{R}/\partial R^2$) to the right hand side of Equation \ref{eq:vrc}. Without $\epsilon c_s$, when $v_{R}$ becomes very small, the LLF flux is reduced to the forward-time-central-difference scheme which is highly oscillatory especially with large additional source terms. To avoid the numerical oscillations, the diffusion timescale ($\Delta R/(\epsilon c_{s})$) should be smaller than the advection timescale ($\Delta R/v_{R}$), which leads to $\epsilon > v_{R}/c_{s}\sim \alpha h$. We found that the results are almost identical when $2\alpha h<\epsilon<1$. It is expected that the diffusion term should not affect the results, since, even with the viscous velocity, the advection term is larger than the diffusion term by a factor of $R/\Delta R$ if $\epsilon\sim \alpha h$. Then, we add the rest of the terms as source terms and $\partial P/\partial R$ is derived using central difference between $i+1$ and $i-1$ grid cells. Treating the centrifugal term and the pressure term as source terms allows us to initialize a perfectly balanced system with zero radial velocity. 

The 1D simulation covers radii from 0.96$\,R_*$ to 100$\,R_*$ with 800 logarithmically spaced cells. For the ghost zones at the inner boundary, the surface density and $\Omega$ are fixed to the stellar values, while the radial velocity is copied from the first active zone. At the outer boundary, the normal zero-gradient boundary conditions are adopted for the surface density and the radial velocity, while $\Omega$ is fixed to the Keplerian value. The remaining disk parameters are identical to those adopted in the 2D simulations.

\section{Results }
\label{sec:results} 

In our 2D simulations we vary the rotation rate of the star with the goal of determining the onset of decretion through the boundary layer. We present the results of three simulations where the rotation rate in the stellar envelope is set to $\Omega_*=0.3$, $0.7$, and $0.8\, \Omega_{\rm b}$. 

\begin{figure}
    \centering
    \includegraphics[width=\linewidth]{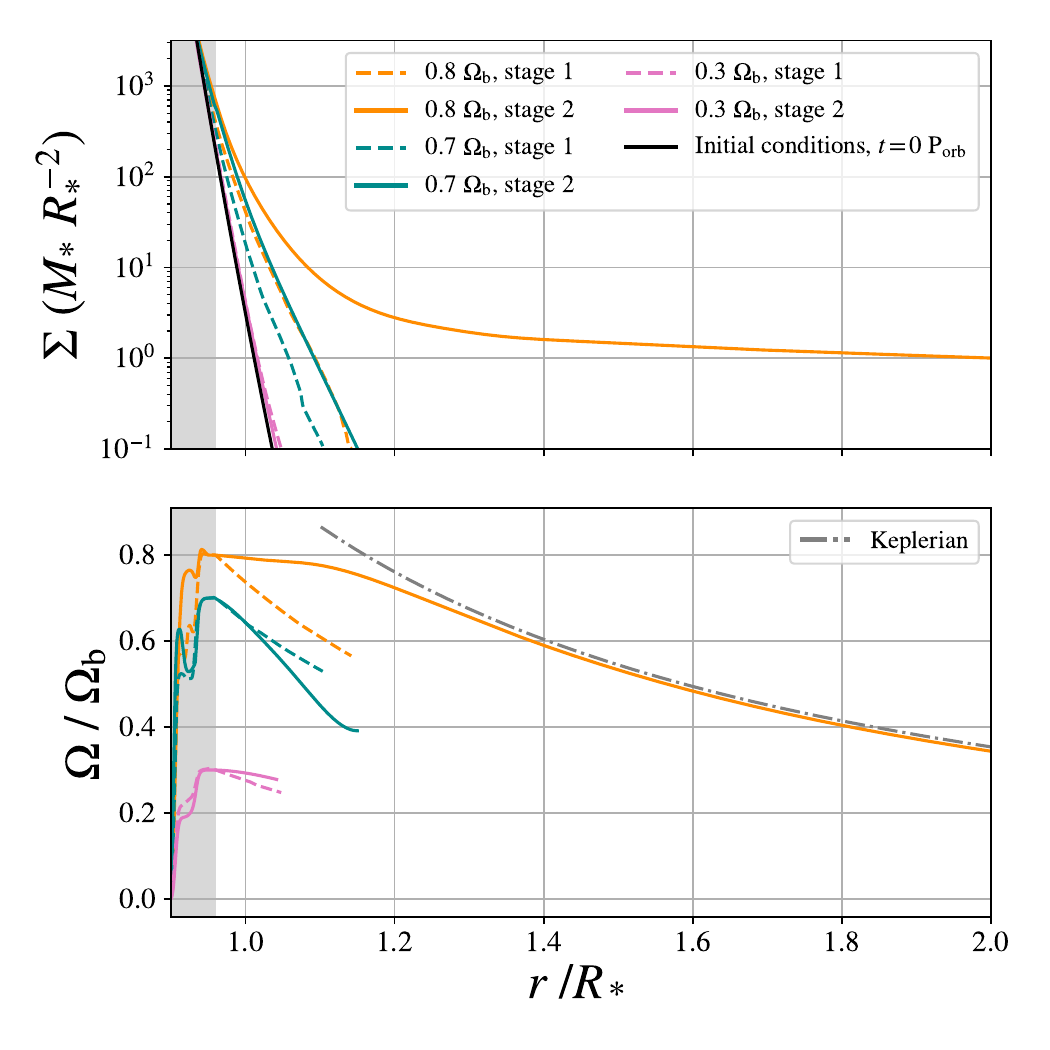}
    \caption{The surface density (upper panel) and the midplane angular velocity (lower panel) as a function of radial distance from the star.  The orange, teal, and pink lines distinguish the three simulations with a stellar rotation of $0.8 \ \Omega_{\rm b}$, $0.7 \ \Omega_{\rm b}$, and $0.3 \ \Omega_{\rm b}$ respectively. The gray region represents $r < 0.96 \ R_*$ where the angular velocity is modified, and is non-physical}. 
    \label{fig:Sigma_Omg_r}
\end{figure}

The top panel of Fig.~\ref{fig:Sigma_Omg_r} shows the surface density as a function of radial distance from the star, for the three simulations at the end of stage~1, (dashed lines) and at the end of stage~2 (solid lines). The density at time $t = 0 \ P_{\rm orb}$ is shown by the black solid line. The surface density, $ \Sigma = \int \rho(r, \theta) \ rd\theta $, is calculated
using the small angle approximation, such that $rd\theta \approx dz$, which is appropriate since the opening half-angle of the disc is $\lesssim 8^{\circ}$ for the inner disc ($r<2 \ R_*$). 

\begin{figure*}
    \centering
    \includegraphics[width=\textwidth]{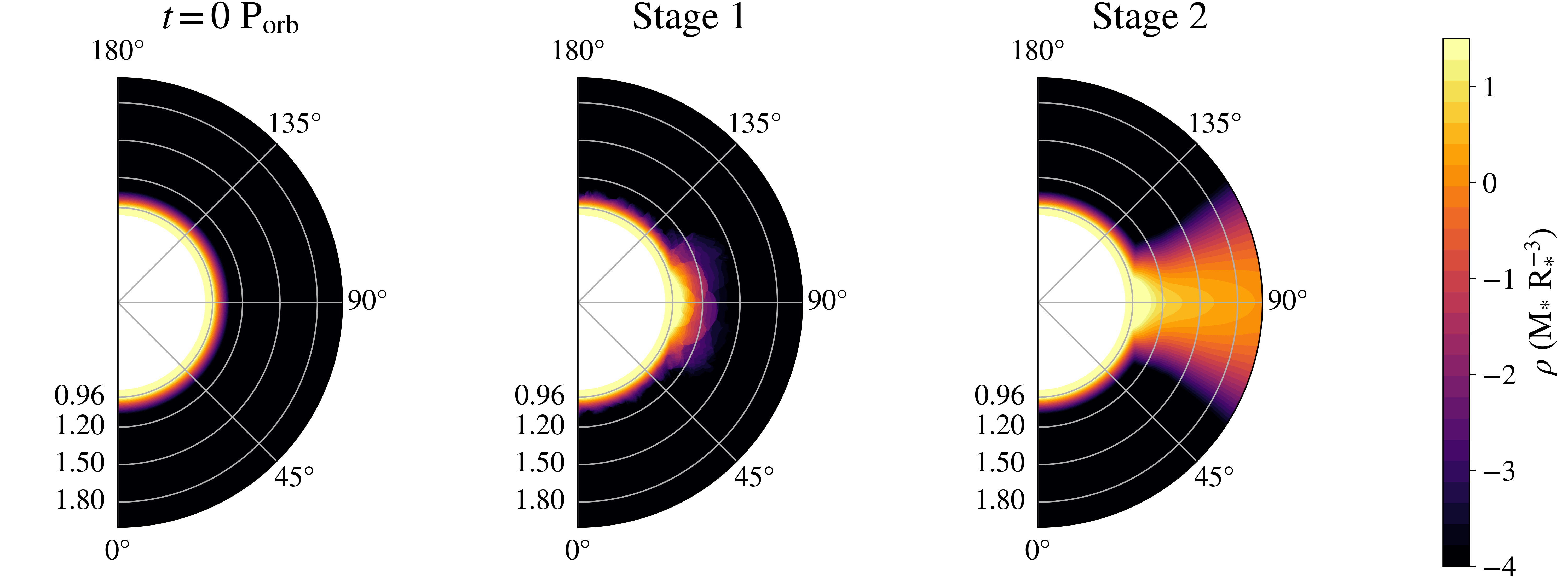}
    \caption{Snapshots of the density for the $\Omega_* = 0.8\,\Omega_{\rm b}$ case at $t = 0 \ P_{\rm orb}$ (left), 
    the end of stage~1 (middle), and the end of stage~2 (right). }
    \label{fig:dens_snap}
\end{figure*}

Recall that in stage~1 we gradually increase the rotation rate of the stellar envelope, and there is no viscosity (see section~\ref{sec:methods}). The gray region represents $r < 0.96 \ R_*$ where the angular velocity is modified, and is non-physical. We only plot values of the surface density greater than $0.1 \, M_*/R_*^2$. As seen by the deviation of the surface density at the end of stage~1 compared to the non-rotating initial conditions, by sufficiently increasing the rotation rate the star develops an equatorial bulge. The stellar envelope becomes more radially extended near the equator and the surface density increases due to rotational support. With larger rotation rates, the equatorial bulge becomes larger. The increase in surface density in the stellar envelope is most evident for the $\Omega_*/\Omega_{\rm b} = 0.7$ and $0.8$ cases, whereas the surface density profile of the $\Omega_*/\Omega_{\rm b} = 0.3$ case does not change significantly as the rotation rate is increased throughout stage~1. 

The bottom panel of Fig.~\ref{fig:Sigma_Omg_r} shows  the midplane angular frequency as a function of radial distance from the star. As described in section~\ref{sec:methods}, the desired rotation rate is achieved at  $r < 0.96 \ R_*$ by the end of stage~1. Beyond $0.96 \, R_*$ the angular velocity decreases with radius. 

Recall that in stage~2, viscosity is turned on for $r \geq 0.96 \ R_*$. In stage~2, the behavior of the envelope significantly differs between the three simulations. In the slowly rotating case with $\Omega_*/\Omega_{\rm b} = 0.3$, including viscosity in the boundary layer does not significantly change the surface density, and the angular velocity profile only slightly flattens out. For $\Omega_*/\Omega_{\rm b} = 0.7$, the boundary layer region becomes more radially extended and experiences an increase in the surface density. The angular velocity in the outer region of the boundary layer (beyond $r \approx 1 \ R_*$) decreases from  its stage~1 values. 
The $\Omega_*/\Omega_{\rm b} = 0.7$ case has enough rotational support to puff up the stellar envelope, but it does not have enough rotational support to fully launch material into the disc. 
An additional factor that may influence the $\Omega_*/\Omega_{\rm b} = 0.7$ case is the interaction of the stellar envelope with the low density material originating from the implementation of a density floor. Recall we implement a density floor of $10^{-8} \ M_*/R_*^3$ to avoid any grid cells approaching zero density. This material develops an inward radial mass flux which may compete with the ability of the boundary layer to launch the disc when the resulting disc's density would be on the order of the density floor. In comparison, for the 0.8 case there is significantly more outflowing material so a small mass flux from the density floor material does not impact the overall behavior in the boundary layer. While the $\Omega_*/\Omega_{\rm b} = 0.3$ and $\Omega_*/\Omega_{\rm b} = 0.8$ cases represent disc behaviors well within the accretion or decretion regimes, respectively, the $\Omega_*/\Omega_{\rm b} = 0.7$ case may represent an intermediary case where the very small radial mass flux of the density floor material becomes significant to the boundary layer behavior.

Once the angular frequency is increased to $\Omega_*/\Omega_{\rm b} = 0.8$, we find that material is able to flow out through the equatorial region of the star, and form a disc. The surface density of the disc region approaches $\Sigma \propto [( R/R_t )^{-1/2} - 1]$, the analytical solution for an expanding decretion disc \citep{Carciofi2008}. However, since the outer disc does not reach a steady state and is still expanding, the truncation radius $R_t$ represents the outer edge of the disc which is not constant in time. 
The angular velocity in the boundary layer increases compared to its stage~1 values but remains a decreasing function with radius. The angular velocity in the disc region becomes very close to the Keplerian value; $\Omega / \Omega_K$ is greater than 0.95 for $r$ between $1.2 \ R_*$ and $2 \ R_*$. The rotation rate is very slightly sub-Keplerian which can be explained by partial pressure support. This is in good agreement with the one dimensional steady state solutions for decretion discs found by \cite{Martinetal2025}. With these features, it's clear that the material that escapes the boundary layer is indeed described by a viscous decretion disc and is not merely an envelope or cloud. 

Fig.~\ref{fig:dens_snap} illustrates the two stages of the simulation, and shows the formation of the decretion disc for the $\Omega_*/\Omega_{\rm b} = 0.8$ case. We plot snapshots of the density at time $t=0 \ P_{\rm orb}$, at the end of stage~1, $t=170 \ P_{\rm orb}$, and at the end of stage~2, $t= 510 \ P_{\rm orb}$. The white region at the origin is outside of the simulation bounds, which extend from $r = 0.9$ to $10 \ R_*$. The leftmost panel shows the initially non-rotating outer layers of the star, as discussed in section~\ref{sec:methods}. The middle panel shows the equatorial bulge that results from increasing the angular velocity within $r < 0.96 \ R_*$. Finally, the rightmost plot shows the decretion disc built up by material flowing outwards through the equatorial region of the boundary layer.  

\begin{figure}
    \centering
    \includegraphics[width=\linewidth]{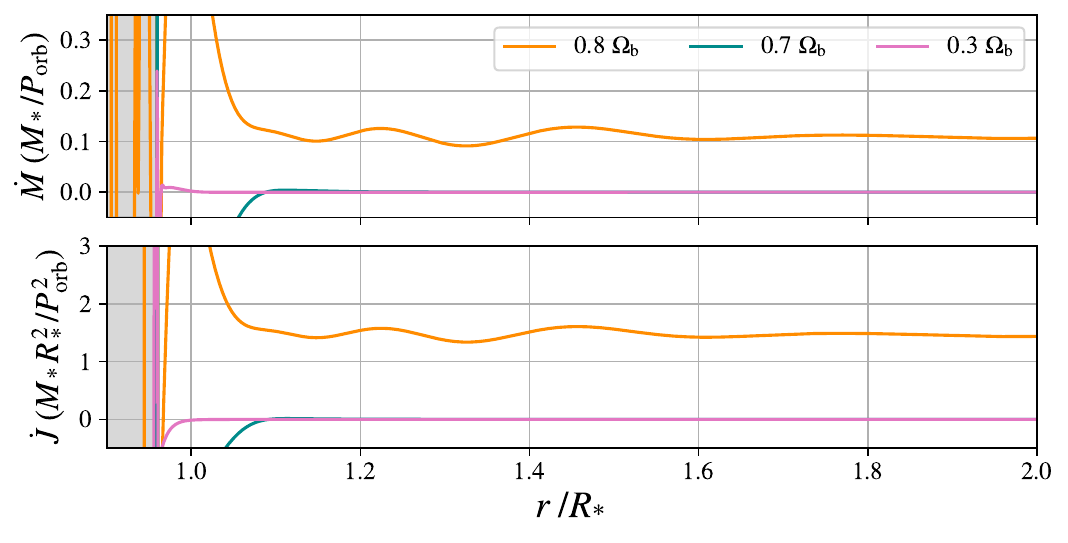}
    \caption{Time averaged mass flux (top panel) and time averaged angular momentum flux (bottom panel). Both quantities are averaged over $100 \ P_{\rm orb}$ at the end of stage~2. Positive values of $\dot{M}$ and $\dot{J}$ indicate decretion and outwards transport of angular momentum. The gray region represents $r < 0.96 \ R_*$ where the angular velocity is modified, and is non-physical.} 
    \label{fig:MdotJdot}
\end{figure}

\begin{figure*}
    \centering
    \includegraphics[width=\textwidth]{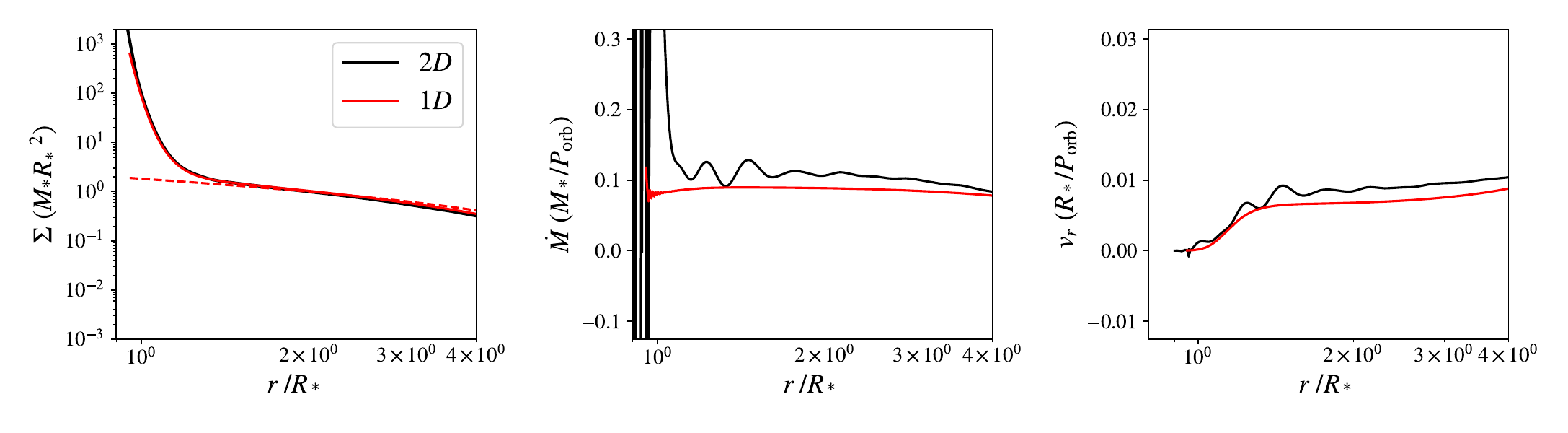}
    \caption{The comparison between the 2D and 1D simulations is shown for surface density (leftmost panel), mass accretion rate (middle panel), and radial velocity ($\dot{M}/(2\pi r\Sigma)$, rightmost panel). The snapshot from the 1D simulation is taken at 510 $P_{\rm orb}$, while quantities are averaged over the final 100 orbits for 2D simulations. In the leftmost panel, the dashed curve shows the analytical solution for the disk surface density with a truncation radius of $8 \ R_*$ \citep{Carciofi2008}. }
    \label{fig:compare}
\end{figure*}

We use the methods outlined in \cite{Dongetal2021} to calculate the time-averaged, radial mass flux $\dot{M}$ and time-averaged, radial angular momentum flux $\dot{J}$ for a two dimensional, axisymmetric setup. The radial mass flux is given by 
\begin{equation}
\label{eqn:Mdot}
    \dot{M}(r_i, \theta_j) = 2 \pi \rho v_r r_i^2 (\cos \theta_{j-1/2} - \cos \theta_{j+1/2})
\end{equation}
for a cell with spherical coordinates $(r_i, \theta_j)$, density $\rho$, and radial velocity $v_r = v_r(r_i, \theta_j)$. In this expression, a positive $\dot{M}$ indicates decretion from the star to the disc. The radial angular momentum flux is given by 
\begin{equation}
\begin{split}
\label{eqn:Jdot}
    \dot{J} = \ & 2 \pi \rho r_i^3 v_r v_{\phi} \bigg[ \frac{1}{2} \big( \theta_{j+1/2} - \theta_{j-1/2} \big) \\
    &- \frac{1}{4} \big( \sin 2 \theta_{j+1/2} - \sin 2 \theta_{j-1/2} \big) \bigg] \\ 
    &- 2 \pi \rho r_i^4 \nu \frac{\Omega_{i+1, \ j} - \Omega_{i, \ j}}{r_{i+1} - r_i} \bigg[ \big( \cos \theta_{j-1/2} - \cos \theta_{j+1/2} \big)  \\
    &- \frac{1}{3} \big( \cos^3 \theta_{j-1/2} - \cos^3 \theta_{j+1/2}\big) \bigg],
\end{split}
\end{equation}
where $v_{\phi}$ is the azimuthal velocity, $\nu$ is the kinematic viscosity. A positive value of $\dot{J}$ indicates that the star is losing angular momentum. 

These quantities are shown in Fig.~\ref{fig:MdotJdot} as a function of the radial distance from the star for the stellar rotation rates of $0.8 \ \Omega_{\rm b}$, $0.7 \ \Omega_{\rm b}$, and $0.3 \ \Omega_{\rm b}$. We average $\dot{M}$ and $\dot{J}$ over the last $100 \ P_{\rm orb}$ at the end of stage~2. For the $\Omega_*/\Omega_{\rm b} = 0.8$ case, both quantities have a flat profile beyond $r \approx 1.2 \ R_*$, indicating this region of the simulation has reached a quasi-steady state. This, together with this region's close adherence to Keplerian rotation (as discussed earlier in this section) provides a convenient way to distinguish the disc from the stellar envelope. Using $\dot{M}$ and $\dot{J}$ we can calculate the constant $j$ given by equation $7$ in \cite{Martinetal2025} to be $j \approx 2$. The parameter $j$ originates from 1D models \citep[][]{Popham&Narayan1991, Paczynski1991, Lee2013} and is related to the truncation radius of the disc for decretion solutions ($j \gg 1$) \citep{Martinetal2025}.
We find positive values of $\dot{M}$ and $\dot{J}$ for the $\Omega_*/\Omega_{\rm b} = 0.8$ case, and negative values very close to zero for $\Omega_*/\Omega_{\rm b} = 0.3$ and $\Omega_*/\Omega_{\rm b} = 0.7$ exterior to the stellar envelope. Values of $\dot{M}$ and $\dot{J}$ close to zero are expected for the absence of material in that region. 

In Figure \ref{fig:compare} we plot the surface density, mass accretion rate, and radial velocity for the 1D model (red lines) alongside the 2D simulations with $\Omega_*/\Omega_{\rm b} = 0.8$ (black lines). We calculate the radial velocity as $v_r = \dot{M}/ (2 \pi r \Sigma)$. All quantities of the 1D model, and the surface density of the 2D simulation are calculated at a time of $510 \ P_{\rm orb}$. The mass accretion rate and the radial velocity for the 2D simulation are averaged over the final 100 orbits of the simulation. 
The surface density of the 1D is scaled to match the inner portions of the 2D simulation's inner disk, for the sake of comparison.
With the same initial conditions, the 1D model reproduces the results from 2D simulations reasonably well.
This agreement is expected, as the equatorial bulge in 2D simulations is already thin at 0.96 $R_*$, where the inner boundary of the 1D simulation is.  

\section{Conclusions}
\label{sec:discussion}

We have presented the first 2D hydrodynamical simulations of the outermost layers of a rapidly rotating star that form a decretion disc at a spin rate that is less than the breakup rate.
For our simulation parameters, ($h_{\rm disc}/r = 0.1$) we find that boundary layer effects can produce a decretion disc for $\Omega_*/\Omega_{\rm b} = 0.8$, but fail to do so for $\Omega_*/\Omega_{\rm b} = 0.7$ and lower. We confirm that the material that escapes the boundary layer has Keplerian rotation and approaches the analytical steady state density profile for a decretion disc, and is therefore well described as a viscous decretion disc.  
We also show that the overall disk behavior is similar to the 1D model. This result aligns with the observed rotation rates of Be stars, and provides a mechanism for the formation of the Be star disk at sub-critical rotation rates. 

This work does not consider the presence of magnetic fields, therefore our results provide a purely hydrodynamical method of building up a Be star disc in the case where magnetic fields are absent. This is of interest given that magnetic fields are undetectable in most Be stars \citep{Neineretal2012}.

This mechanism is compatible with binary formation channels for Be stars, as a massive MS star can acquire a rapid rotation rate through the companion's evolution and resulting mass loss \citep{Shao2014}. However, the launching of the decretion disc in this work is not dependent on a binary companion, as we do not include the influence of a companion in our simulations. 

The rotation rate required for decretion to occur depends on the disc aspect ratio. For circumplanetary disc simulations, \cite{Dongetal2021} found that a larger disc aspect ratio (larger isothermal sound speed) leads to the onset of decretion occurring at lower rotation rates \citep[see also][]{Martin2025b}. This indicates that our results can hold for Be stars with varying disk aspect ratios. However, additional simulations with a larger parameter space are necessary to examine this relationship in more detail. 

We leave the task of more realistic stellar modeling to future work. 
Boundary layer analysis is a common procedure for modeling stellar structure in which the domain is separated into two domains including an interior region which neglects viscosity and a thin, viscous boundary layer \citep[e.g.][]{Espinosa&Rieutord2013}. However, this work does not realistically model the stellar interior. First, the inner boundary of the simulation extends to only $0.9 \ R_*$. We are also limited by the rotational profile we choose interior to $r < 0.96 \ R_*$ which is chosen to avoid numerical instabilities near the inner radial and polar boundaries. The setup used in this work is sufficient for forcing the desired stellar rotation at the inner edge of the boundary layer for the equatorial region of the star. More realistic stellar modeling has been done by incorporating stellar evolution models, radiation hydrodynamics, and magnetohydrodynamics \citep[e.g.][]{Y.-F.Jiangetal2015, Y.-F.Jiangetal2018,  Y.-X.Chenetal2024, ud-Doula2025}. 
Despite the limitations in our methods, as outlined above, this work demonstrates the importance of considering stellar physics and the role of the boundary layer in decretion disc dynamics and evolution.

\section*{Acknowledgements}


We thank Stephen Lubow for useful conversations, and the reviewer for thorough feedback that improved the letter. We acknowledge support from NASA through grant 80NSSC25K0346.

\section*{Data Availability}  

The data underlying this letter will be shared on reasonable request to the corresponding author.



\bibliographystyle{mnras}
\bibliography{bib2025} 








\bsp	
\label{lastpage}
\end{document}